\documentclass[12pt]{article}
\newtheorem{lemma}{Lemma}
\newtheorem{proposition}{Proposition}
\newtheorem{theorem}{Theorem}
\newtheorem{corollary}{Corollary}
\newtheorem{definition}{Definition}
\def\tr{\mathop{\rm Tr}\nolimits}

\def\ci{\mathop{\textrm{i}}\nolimits}
\def\ppsim{\footnotesize\textcircled{\small$\wedge$}}

\catcode`\@=11
\long\def\@makefntext#1{
\protect\noindent \hbox to 3.2pt {\hskip-.9pt
$^{{\ninerm\@thefnmark}}$\hfil}#1\hfill}		

 \def\@makefnmark{\hbox to 0pt{$^{\@thefnmark}$\hss}}  
	
\def\ps@myheadings{\let\@mkboth\@gobbletwo
\def\@oddhead{\hbox{}
\rightmark\hfil\ninerm\thepage}
\def\@oddfoot{}\def\@evenhead{\ninerm\thepage\hfil
\leftmark\hbox{}}\def\@evenfoot{}
\def\sectionmark##1{}\def\subsectionmark##1{}}

\renewenvironment{thebibliography}[1]
 	{\begin{list}{$^{\arabic{enumi}}$}
 	{\usecounter{enumi}\setlength{\parsep}{0pt}
\setlength{\leftmargin 1.25cm}{\rightmargin 0pt}
 	 \setlength{\itemsep}{0pt} \settowidth
 	{\labelwidth}{#1.}\sloppy}}{\end{list}}

\newcounter{itemlistc}
\newcounter{romanlistc}
\newcounter{alphlistc}
\newcounter{arabiclistc}

\def\@citex[#1]#2{\if@filesw\immediate\write\@auxout
 	{\string\citation{#2}}\fi
\def\@citea{}\@cite{\@for\@citeb:=#2\do
 	{\@citea\def\@citea{,}\@ifundefined
 	{b@\@citeb}{{\bf ?}\@warning
 	{Citation `\@citeb' on page \thepage \space undefined}}
 	{\csname b@\@citeb\endcsname}}}{#1}}

\newif\if@cghi
\def\cite{\@cghitrue\@ifnextchar [{\@tempswatrue
 	\@citex}{\@tempswafalse\@citex[]}}
\def\citelow{\@cghifalse\@ifnextchar [{\@tempswatrue
 	\@citex}{\@tempswafalse\@citex[]}}
\def\@cite#1#2{{$\null^{#1}$\if@tempswa\typeout
 	{IJCGA warning: optional citation argument
 	ignored: `#2'} \fi}}

\def\fnt#1#2{\footnotetext{\kern-.3em
 	{$^{\mbox{\sevenrm #1}}$}{#2}}}

 1
 1
 1

\font\ninerm=cmr9

\begin{document}
\title{On the classification of type D spacetimes} 
\author{Joan Josep Ferrando$^1$ and Juan Antonio S\'aez$^2$} 
\date{\empty}

\maketitle
\vspace*{-0.5cm}
\begin{abstract} 
We give a classification of the type D spacetimes based on the invariant differential properties  of the Weyl principal structure. Our classification is established using tensorial invariants of the Weyl tensor and, consequently, besides its intrinsic nature, it is valid for the whole set of the type D metrics and it applies on both, vacuum and non-vacuum solutions. We consider the Cotton-zero type D metrics and we study the classes that are compatible with this condition. The subfamily of spacetimes with constant argument of the Weyl eigenvalue is analyzed in more detail by offering a canonical expression for the metric tensor and by giving a generalization of some results about the non-existence of purely magnetic solutions. The usefulness of these results is illustrated in characterizing and classifying a family of Einstein-Maxwell solutions. Our approach
permits us to give intrinsic and explicit conditions that label every metric, obtaining in this way an operational algorithm to detect them. In particular a characterization of the Reissner-Nordstr\"{o}m metric is accomplished.
\end{abstract}

\vspace*{2mm}
\begin{center}
\end{center}

\vspace*{3mm}
\noindent
$^1$ Departament d'Astronomia i Astrof\'{\i}sica, Universitat
de Val\`encia, E-46100 Burjassot, Val\`encia, Spain.
E-mail: {\tt joan.ferrando@uv.es}\\
$^2$ Departament de Matem\`atica Econ\`omico-Empresarial, Universitat de
Val\`encia, E-46071 Val\`encia, Spain.
E-mail: {\tt juan.a.saez@uv.es}
\newpage

\section{Introduction}

Type D spacetimes have been widely considered in literature and we can point out not only the large number of known families of exact solutions but also the interest of these solutions from the physical point of view. Let us quote, for example, the Schwarszchild or the Kerr metrics which model the exterior gravitational field produced, respectively, by a non-rotating or a rotating spherically symmetric bounded object. Or also the related metrics in the case of a charged object, the Reissner-Nordstr\"{o}m or the Kerr-Newman solutions. However, although some classes of type D metrics have been considered taking into account algebraic properties of the Weyl eigenvalue or differential conditions on the null Weyl principal directions, a classification of the type D solutions involving all the first-order differential properties of the Weyl tensor geometry is a task which has not been totally accomplished yet. In this work we present this classification of the type D metrics and we show the role that it can play in studying geometric properties of known spacetimes, in looking for new solutions of Einstein equations or in offering new elements which allow us to give intrinsic and explicit characterizations of all these spacetimes.

At an algebraic level, a type D Weyl tensor determines a complex scalar invariant, the eigenvalue, and a 2+2 almost-product structure defined by its principal 2--planes. Some classes of type D metrics can be considered by imposing the real or imaginary nature of the Weyl eigenvalue. In this way we find the so called purely electric or purely magnetic spacetimes. The purely electric character often appears as a consequence of usual geometric or physical restrictions\cite{tru}. This is the case of the static type D vacuum spacetimes found by Ehlers and Kundt\cite{ehku}, or the Barnes degenerate perfect fluid solutions with shear-free normal flow\cite{bar}. On the other hand, some restrictions are known on the existence of purely magnetic solutions\cite{mcar,loaa}. A wide bibliography about Weyl-electric and Weyl-magnetic spacetimes can be found in a recent work\cite{fsE} where these concepts have been generalized.

The most usual approaches to look for exact solutions of the Einstein equations work in frames or local coordinates adapted to some outlined direction of the curvature tensor. For example, in the case of perfect fluid solutions or static metrics the 3+1 formalism adapted, respectively, to the fluid flow or to the normal timelike Killing vector can be useful. Sometimes one considers that some of the kinematic coefficients associated with the unitary vector are zero. This means that one is searching for new solutions belonging to a class of metrics that are defined by first-order differential conditions imposed on the curvature tensor. A similar situation appears when local coordinates adapted to the multiple Debever direction are considered when looking for algebraically special solutions. Indeed, if the hypotheses of the generalized Goldberg-Sachs theorem hold, the multiple Debever direction defines a shear-free geodesic null congruence. In this case, or when considering non-diverging or non-twisting restrictions on a Debever direction, we are imposing differential conditions on the Weyl tensor.

It is worth pointing out that the kinematic coefficients associated with a unitary vector completely determine the first-order differential properties of the 1+3 almost-product structure that it defines. Nevertheless, the conditions usually imposed on the two double Debever directions of a type D spacetime do not cover all the differential properties of the principal 2+2 almost-product structure of the Weyl tensor exhaustively. The first goal of this work is to offer a classification of the type D metrics based on all the first-order differential properties of the principal structure, and to reinterpret under this view the usual conditions that can be found in the literature. This classification is not based on the scalar invariants, but on tensorial invariants of the Weyl tensor. These invariants are well adapted to the generic type D metrics, where a Weyl canonical frame is not univocally determined, and where the eigenvalues and the 2+2 principal structure are the only invariants associated with the Weyl tensor. 

The (proper) riemannian almost-product structures have been classified according the invariant decomposition of their structure tensor\cite{nav}, and the classes have been interpreted in terms of the foliation, minimal and umbilical properties\cite{olga}. This classification can be generalized to the spacetime structures by also considering the causal character of the 
planes\cite{cf1}. Almost-product structures have shown their usefulness in studying the underlying geometry of some physical fields. The 1+3 structures are frequently used in relativity and sometimes the properties of a physical field can be expressed in terms of the kinematic properties of a unitary vector\cite{cf2,cf3}. On the other hand, the 2+2 structure associated with a regular solution of Maxwell equations\cite{rai} is a basic concept in building the 'already unified theory' for the electromagnetic field\cite{mis}. It has also allowed a geometric interpretation\cite{cff} of the Teukolsky-Press relations\cite{teu} used in analyzing incident electromagnetic waves on a Kerr black hole.

In General Relativity we can also find almost-product structures attached to the geometric or physical properties of the spacetime. Indeed, some energy contents (for example, in the Einstein-Maxwell or perfect fluid solutions) define underlying structures that restrict, via Einstein equations, the Ricci tensor. On the other hand, the Weyl tensor also defines almost-product structures associated with its principal bivectors depending on the different Petrov types\cite{bel}. These structures determine the Weyl canonical frames\cite{fms}. In the type D case, only the {\it principal structure} is outlined.

Until now we have mentioned two different ways of classifying the type D spacetimes: the first one is strictly algebraic and takes into account the real or imaginary character of the Weyl eigenvalues; the second one, which we will present here, involves differential conditions of the 2+2 principal structure, that is, on the Weyl eigenvectors. Nevertheless, there is a third natural manner to impose restrictions on the type D metrics: to take into account the relative position between the principal 2--planes and the gradient of the Weyl scalar invariants. This is a mixed classification, differential in the eigenvalues and algebraic in the principal structure, which affords 16 different classes of type D metrics. It is worth pointing out the relation that exists between this classification and the two previous ones. On one hand, the real or imaginary nature of the eigenvalues is equivalent to the argument to take particular constant values, and so the purely electric and purely magnetic metrics can be considered as a particular case of argument with zero gradient, a condition that is compatible with 4 of the 16 classes quoted above. On the other hand, in this work we will show that only 16 classes defined by the structure differential properties are compatible with some restrictions of the Ricci tensor that include the vacuum case and, under these hypothesis, these classes coincide precisely with the 16 classes of the mixed classification.

A classification of type D spacetimes taking into account the properties of the 2+2 principal structure shows quite interesting advantages. Indeed, the integration of the static type D vacuum equations using an alternative approach based on the Weyl principal structure has allowed us to complete the results by Ehlers and Kundt\cite{ehku} in order to accomplish an algorithmic and intrinsic identification of the solutions and, in particular, to obtain the equations that define the Schwarzschild spacetime explicitly\cite{fsS}. Moreover, our classification affords a geometric interpretation of the other families of vacuum type D solutions. Starting from this approach two Killing vectors can be determined in terms of Weyl concomitants\cite{fscD}, a result which shows that a commutative bidimensional group of isometries exists . The integration of the type D vacuum equations considering the different classes permits their intrinsic label, as well as a geometric interpretation of the NUT and acceleration parameters\cite{fsDb}. 

In this work we apply our classification to the study of spacetimes with zero Cotton tensor. For them, the Bianchi identities impose the same restrictions on the Weyl tensor as the vacuum condition. We interpret these restrictions in terms of geometric properties of the principal structure and, as we have commented above, we show that the compatible classes can be characterized in terms of the relative position between the gradient of some invariant scalars and the principal 2--planes. From a physical point of view these metrics have two interesting properties. Firstly, the two double Debever directions define shearfree geodesic null congruences and, secondly, the principal structure is maxwellian. This result can be of interest in order to generalize the Teukolsky-Press relations\cite{teu,cff} and their applications to type D non-vacuum solutions.

In order to show the usefulness of this approach in analyzing properties of known metrics, in integrating Einstein equations and in labelling the solutions, here we study the spacetimes with the two properties quoted above for the particular classes with integrable structure. In this case, the spacetime metric turns out to be conformal to a product metric. Then, as a first consequence, we generalize the result by McIntosh et al.\cite{mcar} concerning the non existence of purely magnetic type D vacuum solutions in a double sense: the family of solutions where the new result applies is wider than the vacuum metrics, and the purely magnetic restriction is weakened to an arbitrary constant argument. Moreover, starting from a canonical form we begin on the integration of the Einstein-Maxwell equations for the compatible classes, and we recover the charged counterpart of the A, B, C vacuum metric by Ehlers and Kundt. The integration method at once provides an algorithm to detect the solutions with intrinsic and explicit conditions and, in particular, it offers a characterization of the Reissner-Nordstr\"{o}m spacetime.

The paper is organized as follows. In section 2 we introduce some definitions and notations and we give some results about 2+2 almost-product structures. In section 3 we present the classification of the type D metrics based on the first-order differential properties of the Weyl principal structure, as well as the mixed classification involving the eigenvalues gradient and the principal structure. The Cotton-zero type D metrics are analyzed in section 4, and we show that the principal 2--planes define an umbilical structure and, consequently, we only have 16 compatible classes which coincide precisely with those defined by the mixed classification. The four classes with integrable structure are studied in detail in section 5: we present a canonical form for them and generalize a result about the non-existence of purely magnetic solutions. Finally, in section 6, we apply our results to recover a family of Einstein-Maxwell solutions, to give an operational algorithm to detect them and to explicitly and intrinsically characterize the Reissner-Nordstr\"{o}m spacetime. Some of the results in this paper were communicated without proof at the Spanish Relativity Meeting--96\cite{fscD}.

\section{Spacetime almost-product structures}

On a riemannian manifold $(M,g)$ an almost-product
structure is defined by a p-plane field $V$ and
its orthogonal complement $H$. Let $v$ and $h= g-v$ the respective projectors, and let $Q_v$ be the (2,1)-tensor:
\begin{equation}
Q_v(x,y) = h(\nabla_{vx}vy) \, , \qquad \forall \; \; x,y
\end{equation}
Let us consider the invariant decomposition of $Q_v$ into its antisymmetric part $A_v$ and its symmetric part $S_v \equiv S_v^T + {1 \over p}v \otimes \tr S_v$, where $S_v^T$ is a traceless tensor:
\begin{equation}  \label{Q2}
Q_v = A_v + \frac{1}{p} v \otimes \tr S_v + S_v^T 
\end{equation}
The plane $V$ is foliation if, and only if, $A_v =0$. In this case $Q_v = S_v$ and it coincides with the second fundamental form of the integral manifolds of the foliation $V$\cite{rei}. Moreover $V$ is  minimal, umbilical or geodesic if, and only if, $\tr S_v=0$, $S_v^T =0$ or $S_v  =0$, respectively. Then one can generalize these geometric concepts for plane fields which are not necessarily foliation:
\begin{definition}
A plane field $V$ is said to be geodesic, umbilical or minimal if the symmetric part $S_v$ of its (generalized) second fundamental form $Q_v$ satisfies, respectively, $S_v =0$, $S_v^T =0$ or $\tr S_v  =0$.
\end{definition}

The (proper) riemannian almost-product structures $(V,H)$ have been classified taking into account the invariant decomposition (\ref{Q2}) of the tensors $Q_v$ and $Q_h$ or, equivalently, according with the foliation, minimal, umbilical or geodesic character of each plane\cite{nav,olga}. Some of these properties have also been interpreted in terms of invariance along vector fields\cite{monte}. A generalization for the spacetime structures follows taking into account the causal character of the planes. We will say that a structure is integrable when both planes are foliation and we will say that it is minimal, umbilical or geodesic if both of the planes are so.

This way, on an oriented spacetime $(V_4,g)$ of signature $(-+++)$ we have generically $2^6 =64$ different classes of (almost-product) structures depending on the first-order geometric properties. Nevertheless, when $p=1$, $V$ is always an umbilical foliation and, consequently, only 16 possible classes exist. In this case $Q_v$ and $Q_h$ depend on the kinematic coefficients associated with a unitary vector $u$, and the classes are defined by the vanishing or non-vanishing of the acceleration, rotation, shear and expansion. Elsewhere this kinematical interpretation has been extended to the 2+2 spacetime structures and, as a consequence, the Maxwell-Rainich equations have been expressed in terms of kinematical variables\cite{cf1}.

In order to be used in next sections, we now analyze the spacetime 2+2 almost-product structures in detail by giving the characterization of their properties in terms of their canonical 2--form $U$, and by showing their relation with other usual approches, the Newmann-Penrose and the self-dual formalisms. We also study the change of these properties for a conformal transformation and we summarize some results about maxwellian structures.

\subsection{2+2 structures}
In the case of a 2+2 spacetime structure it is useful to work with the {\it canonical}
unitary 2-form $U$, volume element of the time-like plane $V$. Then, the respective projectors are $v=U^2$ and $h = -(*U)^2$, where $U^2 = U \times U= \tr_{23} U \otimes U$ and $*$ is the Hodge dual operator.
 
The tensors $Q_v$ and $Q_h$ determine the derivatives of the volume elements $U$ and $*U$ by means of 
\begin{equation}\label{UQ}
\begin{array}{l}
\nabla_\alpha U_{\beta \lambda } =(Q_v)_{\alpha \mu,  [\beta } \ {
U^\mu}_{\lambda]} + (Q_h)_{\alpha [\beta ,}^{\ \ \ \ \mu} \ U_{\lambda] \mu}
\\[3mm]
\nabla_\alpha *U_{\beta \lambda} =(Q_h)_{\alpha \mu , [\beta } \
{*U^\mu}_{\lambda]} + (Q_v)_{\alpha [\beta ,}^{\ \ \ \ \mu} \ *U_{\lambda] \mu}
\end{array}
\end{equation}
Then, if we denote $\delta = - \tr \nabla$, a straightforward calculation leads to: 

\begin{equation} \label{UtrQ}
\delta U = i({\rm \tr} S_h)U - 2(U,A_v) \qquad   
\delta *U = i({\rm \tr} S_v)*U - 2(*U, A_h)
\end{equation}
where $2(U, A_v)^{\mu} = U^{\alpha \beta}{(A_v)_{\alpha \beta}}^{ \mu}  
$. So, the minimal and the foliation character of the planes can be stated in terms of the projections of $\delta U$ and $\delta *U$ onto $V$ and $H$. On the other hand, let us consider:
 
\begin{equation}\label{umbilical1} 
G_{\perp} = U \otimes U - *U \otimes *U + G ; \hspace*{1cm}
\eta_{\perp} = U \stackrel{\sim}{\otimes} *U + \eta 
\end{equation} 
where $\eta$ is the metric volume element of the spacetime, $G=\frac{1}{2} g {\ppsim} g$ is the metric on the 2--forms space,  and ${\ppsim}$ denotes the double-forms exterior product, $(A {\ppsim} B)_{\alpha \beta \mu \nu} = A_{\alpha \mu} B_{\beta \nu} + A_{\beta \nu} B_{\alpha \mu} - A_{\alpha \nu} B_{\beta \mu} - A_{\beta \mu} B_{\alpha \nu}$. The
tensors (\ref{umbilical1}) satisfy $G_{\perp} (U) =G_{\perp} (*U) =0$,
$\eta_{\perp} (U) = \eta_{\perp} (*U) =0$ and they can be calculated as
 
\begin{equation} \label{umbilical2} 
G_{\perp} = v {\ppsim} h, \qquad  \eta_{\perp} = U {\ppsim} *U 
\end{equation}
Then, from expressions (\ref{UQ}) and (\ref{UtrQ}) we get 
\begin{eqnarray} 
(2 \nabla U - K )_{\lambda \alpha \beta} = (S_{v}^{T})_{\lambda \mu, 
[\alpha} \ {U^{\mu}}  _{\beta ]} + (S_{h}^{T})_{\lambda \mu, 
[\alpha} \ {*U^{\mu}}_{\beta]}, \label{umbilical3} \\[2mm]
K \equiv  i(\delta U) G_{\perp} - i(\delta *U) \eta_{\perp} \label{umbilical4}
\end{eqnarray}
and so, the umbilicity of each plane is equivalent to the vanishing of the respective projections of the first member of (\ref{umbilical3}). We summarize these results in the following lemma:

\begin{lemma} \label{lem-2+2}
 Let $(V,H)$ be a 2+2 almost-product structure and
let $U$ be its canonical 2-form. Then, the following conditions hold:
\begin{enumerate} 

\item $V$ (resp. $H$) is foliation $\Longleftrightarrow$ $i(\delta 
U)*U = 0$
(resp. $i(\delta *U)U = 0$).

\item $V$ (resp. $H$) is minimal $\Longleftrightarrow$ $i(\delta 
*U)*U = 0$
(resp. $i(\delta U)U = 0$).

\item $V$ is umbilical $\Longleftrightarrow$ 
$U \times \{2 \nabla U - [i(\delta U) G_{\perp} - i(\delta *U) 
\eta_{\perp}]\} =0$\\
$\quad H$ is umbilical $\Longleftrightarrow$  $*U \times \{2 \nabla U - [i(\delta U) G_{\perp} - i(\delta *U) \eta_{\perp} ]\}=0$
\end{enumerate} 
\end{lemma}

A 2+2 structure is also determined by the two null directions $l_{\underline{+}}$ on the plane $V$. A family of complex null bases $\{ l_+ , l_- , m, \bar{m} \} $ exists such that $U=l_{-} \wedge l_{+}$. This family is fixed up to change $l_{\pm} \hookrightarrow e^{\pm \phi} l_{\pm}$, $m  \hookrightarrow e^{{\rm {i}} \theta} m$. Then, conditions of lemma \ref{lem-2+2} can be interpreted in terms of the Newman-Penrose coefficients\cite{kra} as

\begin{lemma} \label{lem-NP}
Let $U= l_- \wedge l_+ $ be the canonical 2--form of a 2+2 structure. It holds:
\begin{enumerate} 
\item The plane $V$ is umbilical iff  $\kappa=0=\nu$. 
\item The plane $H$ is umbilical iff  $\lambda =0=\sigma$.  
\item The plane $V$ is minimal iff $\bar{\pi} = \tau$.
\item The plane $H$ is minimal iff $\rho+ \bar{\rho}=0= \mu+ \bar{\mu}$. 
\item The plane $V$ is a foliation iff $\bar{\pi} = - \tau$.
\item The plane $H$ is a foliation iff $\rho- \bar{\rho} =0= \mu- \bar{\mu} $. 
\end{enumerate} 
\end{lemma} 
Taking into account the significance of the NP coefficients\cite{kra} this lemma implies that the umbilical nature of a 2+2 structure means that its principal directions $l_{\pm}$ define shear-free geodesic null congruences. The minimal or foliation character of the spacelike 2--plane have also a kinematical interpretation and state, respectively, that both principal directions are expansion-free or vorticity-free. Elsewhere\cite{cf1} all the geometric properties have been interpreted in terms of kinematic coefficients associated with every direction in a 2--plane (not only the null ones) with respect to the other 2--plane.

When both planes have a specific differential property, it is more convenient to introduce the self-dual unitary 2--form 
${\cal U} \equiv \frac{1}{\sqrt{2}} (U - {\rm {i}} *U )$ associated with $U$. We have
\begin{equation} \label{calU}
\begin{array}{l}
2 \mbox{Re} [ i(\delta {\cal U}) {\cal U} ] =
i(\delta U) U - i(\delta *U)*U \equiv \Phi(U) \\[3mm]
 2  \mbox{Im} [ i (\delta {\cal U}) {\cal U} ] =
- i(\delta U)*U -  i(\delta *U) U \equiv \Psi(U)
\end{array}
\end{equation}
So, the complex 1--form $\delta {\cal U}$ collects the information about the minimal and foliation  character  of the structure. On the other hand, if ${\cal G} = \frac{1}{2} ( G - {\rm i} \eta )$ is the metric on the self-dual 2--forms space, and ${\cal K} \equiv  \frac{1}{\sqrt{2}} (K - {\rm {i}} *K )$ is the self-dual 2-form associated to the vector valued 2--form $K$ given in (\ref{umbilical4}), we have
\begin{equation} \label{calUu}
{\cal K} = i(\delta {\cal U}) [ {\cal U} \otimes {\cal U} + {\cal G} ]  
\end{equation}
Consequently, from lemma \ref{lem-2+2} and equations (\ref{calU}) and (\ref{calUu}), we have: 

\begin{lemma}  \label{lem-S-D}
Let us consider the 2+2 structure defined by ${\cal U}=\frac{1}{\sqrt{2}} (U - i *U)$. It holds: 
\begin{enumerate} 
\item The structure is minimal if, and only if, $\mbox{Re} [ i(\delta {\cal U}) {\cal U} ]=0 $.
\item The structure is integrable if, and only if, $\mbox{Im} [ i(\delta {\cal U}) {\cal U} ]=0$.
\item The structure is umbilical, if, and only if, $\nabla {\cal U} = i(\delta {\cal U}) [ {\cal U} \otimes {\cal U} + {\cal G} ] $.
\end{enumerate} 
\end{lemma} 
  
\subsection{Almost-product structures and conformal transformations}

If $(V,H)$ is a p+q almost-product structure for a metric $g$, then $(V,H)$ is also an almost-product structure for every conformal metric $\hat{g} = e^{2 \lambda} g$, and the projectors are related by the conformal factor: if $g=v+h$, then $\hat{g} = \hat{v} + \hat{h}$, where  
$\hat{v} =  e^{2 \lambda}v$, $\hat{h} = e^{2 \lambda} h$. The generalized second fundamental form change as 
\begin{equation}\label{Q-con}
Q_{\hat{v}} = e^{2\lambda} \  ( Q_v - v \otimes h(\mbox{d} \lambda) )
\end{equation}
So, the foliation and the umbilical character are conformal invariants, but the minimal character is not. Indeed, taking the trace of the expression above, we have 
\begin{equation}
\tr Q_{\hat{v}} = \tr Q_v - p h(\mbox{d} \lambda) 
\end{equation}
These expressions inmediately lead to the following result.

\begin{lemma}\label{lem-con}
Let $(V,H)$ be a p+q almost-product structure for a metric $g=v+h$. The structure $(V,H)$ is minimal for a conformal metric $\hat{g} = e^{2 \lambda} g$ if, and only if, 
\begin{equation} \label{con}
\frac{1}{p} \tr{Q_v} + \frac{1}{q} \tr Q_h = \mbox{\rm d} \lambda
\end{equation}
\end{lemma}
If $p=q$ (as happens for the spacetime 2+2 structures), we conclude that the necessary and sufficient condition for a structure to be minimal for a conformal metric is the sum of the traces of the second fundamental forms to be a closed 1-form, $\mbox{d}(\tr Q_v + \tr Q_h)=0$. Thus, taking into account (\ref{UtrQ}) and the expression (\ref{calU}) for $\Phi(U)$, lemma \ref{lem-con} can be stated for the 2+2 case as

\begin{lemma}\label{lem-conU}
Let $U$ be the canonical 2--form of a 2+2 structure for the spacetime metric $g$. The structure is minimal for a conformal metric if, and only if, $\mbox{\rm d} \Phi(U) = 0$. More precisely, when this condition hold, let $\lambda$ be such that $2 \mbox{\rm d} \lambda = \Phi(U)$. Then, the structure is minimal for the conformal metric $\hat{g} = e^{2 \lambda} g$. 
\end{lemma}

The most degenerated class of almost-product structures are the product ones, which means, those that satisfy $Q_v =0 = Q_h$. A metric that admits a product structure is called a product metric. 
Then, and only then, local coordinates
$(x^{ A}, x^{i})$, $A=0,1, \ i=2,3$, exist such that
$\tilde{g}=\sigma^{-}+\sigma^{+}$,
 being $\sigma^{-}=\sigma^{-}_{AB}(x^{C}) \mbox{d}x^{A}
\mbox{d}x^{B}$ and $\sigma^{+}=\sigma^{+}_{ij}(x^{k}) \mbox{d}x^{i} \mbox{d}x^{j}$ bidimensional metrics, hyperbolic and elliptic, respectively. Then, if  $\tilde{g}$ is a 2+2 product metric and  $g = e^{-2 \lambda} \tilde{g}$, lemma \ref{lem-conU} and expression (\ref{Q-con})
lead to the following result.

\begin{lemma}  \label{lem-product}
The necessary and sufficient condition for a metric $g$ to be conformal to a product metric $\tilde{g}$, is that an integrable and umbilical almost-product structure $U$ exists such that $\mbox{\rm d} \Phi(U) = 0$. More precisely, if $2 \mbox{\rm d} \lambda = \Phi(U)$, then 
$\tilde{g} = e^{2 \lambda} g$ is a product metric. 
\end{lemma}

\subsection{Maxwellian structures}

A regular 2-form $F$ takes the canonical expression $F = e^{\phi}[\cos \psi U + \sin \psi *U]$, where $U$ defines the 2+2 associated structure, $\phi$ is the {\it energetic index} and $\psi$ is the {\it Rainich index}. When $F$ is solution of the source-free Maxwell equations, $\delta F =0$, $\delta *F =0$, one says that $U$ defines a {\it maxwellian structure}. In terms of the canonical elements $(U,\phi,\psi)$, Maxwell equations write\cite{rai,cff}:
\begin{eqnarray} 
\mbox{d} \phi = \Phi(U) \equiv i(\delta U) U - i(\delta *U)*U  \label{max1a} \\[2mm]
\mbox{d} \psi = \Psi(U) \equiv - i(\delta U)*U -  i(\delta *U) U  \label{max1b}
\end{eqnarray} 
Then, from (\ref{max1a}) and (\ref{max1b}) the Rainich theorem\cite{rai} follows:

\begin{lemma}
A unitary 2-form $U$ defines a maxwellian structure if, and only if, it satisfies:
\begin{equation}\label{max2} 
\mbox{\rm d} \Phi(U)  = 0 ; \qquad  \qquad  \qquad
\mbox{\rm d} \Psi(U)  = 0 
\end{equation}
\end{lemma}
The Maxwell-Rainich equations (\ref{max1a}) and (\ref{max1b}) have a simple expression in the self-dual formalism. Indeed, the self-dual $2$--form ${\cal F}= \frac{1}{\sqrt{2}} (F- \ci *F)$ writes ${\cal F} = e^{\phi + \mbox{\rm i} \psi} \ {\cal U}$. Then, from Maxwell equations, $\delta {\cal F}=0$, and taking into account that $2 \, {\cal U}^{\,2} = g$, 
\begin{equation}
\mbox{d} (\phi + \mbox{\rm i} \psi) = 2 i(\delta {\cal U}) {\cal U}
\end{equation}
This last equation is equivalent to (\ref{max1a}) and (\ref{max1b}) if we take into account (\ref{calU}). Moreover, from here we recover the complex version of (\ref{max2}) easily
\begin{equation}
\mbox{d} i(\delta {\cal U}) {\cal U} = 0
\end{equation}

\section{Classifying type D spacetimes}

The self--dual Weyl tensor ${\cal W} =\frac{1}{2} (W - {\rm{i}} *W)$ of a type D spacetime takes the canonical expression\cite{fms} 
\begin{equation} \label{cano}
{\cal W} = 3 \alpha \ {\cal U} \otimes {\cal U} + \alpha   \  {\cal G}
\end{equation}
where $\alpha = - \frac{{\rm \tr}{\cal W}^3}{{\rm \tr}{\cal W}^2}$ is the double eigenvalue and ${\cal U}$ is the self-dual principal 2--form. This principal 2--form defines a 2+2 almost-product structure which is called the {\it principal structure} of a type D spacetime. In terms of the canonical 2--form $U$ of the principal structure the self-dual 2--form ${\cal U}$ writes ${\cal U}= \frac{1}{\sqrt{2}} (U- \ci *U)$. So, at the algebraic level, a type D Weyl tensor only determines the complex scalar $\alpha$ and the principal structure $U$. Consequently, any generic classification of the type D metrics must depend on these invariants associated with the Weyl tensor. 

The families of purely electric or purely magnetic type D spacetimes are defined, at first glance, by means of alternative conditions, namely, the nullity of the magnetic or the electric Weyl fields associated with an observer $u$. But, actually, they admit a simple intrinsic characterization in terms of the Weyl scalar invariant: the eigenvalue is real or 
imaginary\cite{mcar}. In spite of these strong conditions, the family of Weyl-electric type D spacetimes contains quite interesting solutions. We can quote, for example, the static vacuum metrics\cite{ehku} or the degenerate perfect fluids with shearfree normal flow\cite{bar}. All the type D silent universes are also known\cite{marc,sze} as well as other families of purely electric type D perfect fluid solutions\cite{co,baro}. Nevertheless, few Weyl-magnetic type D solutions have been found\cite{loaa}, and some restrictions about their existence are known. Indeed, there are not vacuum metrics with purely magnetic type D Weyl tensor\cite{mcar}. The classification that we present below allows us to give an extension of this result in section 5. On the other hand, the generalization of the purely electric or magnetic concepts to the spacelike or null directions does not afford new classes in the type D case\cite{fsE}.

But the purely electric or magnetic properties define very narrow subsets of the generic type D metrics because they impose one of the two real scalar invariants to be zero. The large family of known solutions of the Einstein equation recommends us to consider other classifications, based on less restrictive properties, which afford new intrinsic elements that increase the knowledge of the metrics and permit their explicit characterization. Besides the {\it intrinsic} nature, the classification must be {\it generic}, that is, valid for the whole set of the type D metrics. Consequently, it will be independent of the energy content and it will have to be built on the intrinsic geometry associated with a type D Weyl tensor.  

The first classification that we propose is based on the geometric properties of the principal 2--planes, that is, it is induced by the geometric classification of the principal structure. Every principal 2--plane can be submitted or not to three properties, so $2^6 = 64$ classes can be considered.
\begin{definition}
Taking into account the foliation, minimal or umbilical character of each principal 2--plane we distinguish $64$ different classes of type D spacetimes.

We denote the classes as D$^{pqr}_{lmn}$, where the superscripts $p,q,r$ take the value $0$ if the time-like principal plane is, respectively, a foliation, a minimal or an umbilical distribution, and they take the value $1$ otherwise. In the same way, the subscripts $l, m , n$ collect the foliation, minimal or umbilical nature of the space-like plane.
\end{definition} 

The most degenerated class that we can consider is D$^{000}_{000}$ which corresponds to a type D product metric, and the most regular one is D$^{111}_{111}$ which means that neither $V$ nor $H$ are foliation, minimal or umbilical distributions. We will put a dot in place of a fixed script (1 or 0) to indicate the set of metrics that cover both possibilities. So, for example, the metrics of type D$^{111}_{11 \, \cdot}$ are the union of the classes D$^{111}_{111}$ and D$^{111}_{110}$; or a metric is of type D$^{0 \, \cdot \, \cdot}_{\, \cdot \, \cdot \, \cdot}$ if the timelike 2--plane is a foliation. 

Taking into account lemma \ref{lem-2+2}, every class is defined by means of first-order differential equations imposed on the canonical 2--form $U$. On the other hand, $U$ can be written explicitly in terms of the Weyl tensor\cite{fms} and, consequently, every class admits an intrinsic and explicit characterization.

The above classification depends on the derivatives of the principal 2--form $U$. An alternative classification at first order in the Weyl eigenvalues can also be considered by taking into account the four 1-forms defined by the principal 2--planes and the gradient of the modulus and the argument of the eigenvalue. So, we will have $2^4 = 16$ classes.

\begin{definition}
Let $\alpha = e^{\frac{3}{2}( \rho +  {\rm{i}} \theta)}$ be the Weyl eigenvalue. Taking into account the relative position between the gradients $d \theta, d \rho$ and each principal 2--plane we distinguish $16$ different classes of type D spacetimes.

We denote the classes D$[pq,rs]$ where $p,q,r,s$ take the values $0$ or $1$ to indicate, respectively, that  one of the 1-forms $U(d \theta) , U(d \rho), *U(d \theta) , *U(d \rho)$ is zero or non-zero.
\end{definition} 

The most degenerated class D$[00;00]$ is ocuped by the type D metrics with constant eigenvalues, and the most general one 
D$[11;11]$ by those type D spacetimes for which both, the modulus and the argument of the Weyl eigenvalue, have non zero projection onto the principal planes. As above, a dot means that a condition is not fixed. So, for example, we write D$[0 \, \cdot \, ; \, \cdot \, \cdot]$ to indicate the type D metrics for which the argument of the eigenvalues have zero projection onto the timelike principal 2--plane. 

The type D metrics with constant modulus, $d\rho=0$, correspond to the classes $D[\, \cdot 0; \, \cdot 0]$, and those with constant angument, $d\theta =0$, are the metrics of type $D[0 \, \cdot ;0\, \cdot ]$. This last family contains the Weyl-electric and the Weyl-magnetic spacetimes because a real or imaginary eigenvalue means that the argument takes the constant value $0, \pi$ or $\pi/2, 3\pi/2$, respectively. 

In the next section we will show the marked relation between the two classifications given in definitions 2 and 3 when some usual restrictions are imposed on the Ricci tensor.

\section{Type D metrics with zero Cotton tensor}

The spacetime Cotton tensor $P$ is a vector valued 2--form which depends on the Ricci tensor as
\begin{equation} \label{cotton}
P_{\mu \nu , \beta} \equiv \nabla_{[\mu} Q_{\nu] \beta} \ , 
\qquad \quad 2Q \equiv Ric - \frac{1}{6} (\tr Ric) g
\end{equation}
The Bianchi identities equal the Cotton tensor with the divergence of the Weyl tensor. Indeed, if ${\cal W}$ is the self-dual Weyl tensor and ${\cal P} = \frac{1}{2} ( P - {\rm i} *P)$ is the self-dual 2--form associated with the Cotton tensor, Bianchi identities write
\begin{equation} \label{bianchi1} 
{\cal P} = - \delta {\cal W}   
\end{equation}
So, the vanishing of the Cotton tensor is equivalent to the Weyl tensor to be divergence free, $\delta {\cal W}=0$. Taking into account the canonical expression of a type D Weyl tensor (\ref{cano}), a straightforward calculation leads to the following

\begin{proposition} \label{prop-cotton0}
Let ${\cal U}$ and $\alpha=- \frac{\tr{\cal W}^3}{\tr{\cal W}^2} $  be the principal  2--form and the double eigenvalue of a type D Weyl tensor. Then, the spacetime Cotton tensor is zero if, and only if,
\begin{equation} \label{divcero}
\nabla {\cal U} = i(\delta {\cal U}) [{\cal U} \otimes {\cal U} + {\cal G} ] \ ; \ \quad
i (\delta{\cal U} ) {\cal U} = \frac{1}{3} \ \mbox{\rm d} \ln{\alpha}
\end{equation}  
\end{proposition} 
From the results of the previous section, we know that the first condition means that the principal structure is umbilical, that is, the principal directions are shear free null geodesics accordingly to the Goldberg-Sachs theorem. Consequently, every type D spacetime with zero Cotton tensor is of type D$^{\cdot \cdot 0}_{\cdot \cdot 0}$. The second equation in (\ref{divcero}) shows that the principal structure is maxwellian and the electromagnetic invariant scalars depend on the Weyl eigenvalue. If we take the real and the imaginary parts of this equation and write $ \rho +  {\rm i} \theta = \frac{2}{3} \ln \alpha$, we get 
\begin{equation} \label{divcero1} 
\Phi(U) =  \mbox{d} \rho \ ; \qquad \qquad \Psi(U) = \mbox{d} \theta 
\end{equation}
So the modulus and the argument of the Weyl eigenvalue govern, respectively, the minimal and the foliation character of the principal planes. This relation establishes a bijection between the classes of the two classifications that we have presented. More precisely, we have: 

\begin{theorem} \label{th-cotton}
Every type D spacetime with zero Cotton tensor is of type D$^{\cdot \cdot 0}_{\cdot \cdot 0}$. Moreover, it is of class D$^{pq0}_{lm0}$ if, and only if, it is of class 
D$[lm,pq]$
\end{theorem}
So we have just 16 classes of type D spacetimes with zero Cotton tensor 
and each one is characterized by the vanishing or not of the projections of the gradient of the Weyl eigenvalue onto the principal planes. 
The second condition in (\ref{divcero}) implies that a solution of the Maxwell equations exists that has ${\cal U}$ as its associated structure. Then, taking into account the results of the subsection 2.3, it holds:

\begin{proposition} \label{pro-cotton}
The principal structure of a type D spacetime with zero Cotton tensor is maxwellian. More precisely, if ${\cal U}$ and $\alpha=- \frac{\tr{\cal W}^3}{\tr{\cal W}^2} $ are the principal 2--form and the double eigenvalue of the Weyl tensor, the self--dual 2--form 
\begin{equation} \label{maxwelld}
{\cal F}_M = \alpha^{\frac{2}{3}} {\cal U}
\end{equation}
is a solution of the source-free Maxwell equations, $ \delta {\cal F}_M = 0$. 
\end{proposition} 
It is worth pointing out that the family of type D metrics admitting a conformal Killing-Yano tensor attached to its principal structure are those of type D(M)$^{\cdot \cdot 0}_{\cdot \cdot 0}$ \cite{fsR}. This family of spacetimes includes the Cotton-zero type D metrics and it has been intrinsically characterized elsewhere in terms of the Weyl and Cotton tensors \cite{cfs}. 

In the following D(M) denotes the type D spacetimes with maxwellian principal structure, and D(M)$^{pqr}_{lmn}$ expresses the type D(M) spacetimes of class D$^{pqr}_{lmn}$. With this notation, from theorem \ref{th-cotton} and proposition \ref{pro-cotton} it follows: {\it every type D spacetime with zero Cotton tensor is of type} $D(M)^{\cdot \cdot 0}_{\cdot \cdot 0}$. 

It is worth pointing out that the family of type D metrics admitting a conformal Killing tensor attached to its principal structure contains those of type D(M)$^{\cdot \cdot 0}_{\cdot \cdot 0}$. This result has been shown elsewhere\cite{cfs} and this family of spacetimes, which includes the Cotton-zero type D metrics, has been intrinsically characterized in terms of the Weyl and Cotton tensors.

The results of this section have been used elsewhere in offering a new approach to the type D vacuum solutions\cite{fsDb}. An integration of the Einstein vacuum equations based on the classification given above permits the explicit and intrinsic labeling of the solutions as well as to put over interesting geometric properties of these spacetimes.

\section{Some results about type D(M)$^{0 \cdot 0}_{0 \cdot 0}$ spacetimes}

Now, in this section, we restrict our study to the type D metrics with maxwellian, integrable and umbilical structure, that is, those of type 
D(M)$^{0 \cdot 0}_{0 \cdot 0}$. We can easily obtain a canonical form for these metrics. Indeed, lemma \ref{lem-product} states that the metric is conformal to a product one with a conformal factor determined by the potential of the closed 1--form $\Phi(U)$. More precisely, the metric can be written
\begin{equation} \label{canonica} 
g = \frac{1}{\Omega^2} \left[ 
\sigma^{-}_{AB} (x^C) \ d x^A d x^B +  \sigma^{+}_{ij} (x^k) \ d x^i d x^j \right]
\end{equation}
where $\Omega$ satisfies
\begin{equation}  \label{canonicfactor}
2 \, \mbox{d} \ln \Omega = \Phi(U) \equiv i(\delta U) U - i(\delta *U)*U 
\end{equation}

Conversely, we can analyze the Petrov type of the metric (\ref{canonica}) by studying a product metric $\tilde{g} = \sigma^{-}+\sigma^{+}$. Let $X_{-} $ and $X_{+} $ be the gaussian curvatures of the arbitrary bidimensional metrics, $\sigma^{-}$ and $\sigma^{+}$, hyperbolic and elliptic respectively. The Gauss-Codazzi equations show that the Riemann and the Ricci tensors of $\tilde{g}$ are 
\begin{equation} \label{producto1}
 Riem (\tilde{g}) = \frac{1}{2} X_{-}  \sigma^{-} {\ppsim} \sigma^{-} +
\frac{1}{2} X_{+}  \sigma^{+} {\ppsim} \sigma^{+} ; \ \ \ \ 
 Ric(\tilde{g}) = X_{-}  \sigma^{-} + X_{+}  \sigma^{+}
\end{equation} 
So, the Weyl tensor of a product metric is Petrov type O precisely when $X_{-} +X_{+}  =0$, and then both curvatures are constant. On the other hand, when $X_{-} +X_{+}  \not= 0$, the spacetime is type D. Moreover $U$ determines the principal structure and the double eigenvalue is given by 
\begin{equation} \label{producto2} 
\tilde{\alpha}=- \frac{1}{6}(X_{-}  + X_{+} ),
\end{equation} 
So, we have
\begin{lemma} \label{lemma-proD}
Every 2+2 product metric $\sigma^{-}+\sigma^{+}$ is of type D (or O) with real eigenvalues, and the double eigenvalue is given by (\ref{producto2}), where $X_{-} $ and $X_{+} $ are the gaussian curvatures of $\sigma^{-}$ and $\sigma^{+}$, respectively. Moreover, it is of type O if, and only if, $X_{-} =-X_{+}  = constant$.
\end{lemma}

A conformal transformation $\tilde{g} = \Omega^2 g$ preserves the Petrov type and the Weyl eigenvalues change as $\tilde{\alpha} = \Omega^{-2} \ \alpha$. Consequently, from equation (\ref{canonicfactor}) and taking into account lemmas \ref{lem-2+2} and \ref{lemma-proD}, we can conclude: 

\begin{proposition} \label{pro-inte1}
A spacetime is of type D(M)$^{0 \cdot 0}_{0 \cdot 0}$ if, and only if, there exist local coordinates such that the metric $g$ takes the expression (\ref{canonica}) with $X_{-} +X_{+}  \neq 0$, where $X_{-} $ and $X_{+} $ are the gaussian curvatures of 
$\sigma^{-}$ and $\sigma^{+}$, respectively. Moreover, it is of class D$^{010}_{010}$,  D$^{000}_{010}$, D$^{010}_{000}$ or 
D$^{000}_{000}$ if, and only if, $\sigma^{-}(d \Omega) \neq 0 \neq \sigma^{+}(d \Omega)$, $\sigma^{+}(d \Omega)=0 \neq \sigma^{-}(d \Omega)$, $\sigma^{+}(d \Omega) \neq 0 = \sigma^{-}(d \Omega)$ or $d \Omega =0$, respectively. 
\end{proposition}
Furthermore, taking into account the expressions (\ref{producto1}) for the Ricci and (\ref{producto2}) for the eigenvalue of a product metric, and considering the change of these metric concomitants for a conformal transformation, we can state:

\begin{proposition} \label{pro-inte2}
The Weyl eigenvalue of the metric (\ref{canonica}) is real and it is  given by
\begin{equation}  
\alpha = - \frac{1}{6} \Omega^2 (X_{-} + X_{+} )  \label{valorpropio} 
\end{equation}
The Ricci tensor of this metric is
\begin{equation}
Ric(g)=  {{2}\over{\Omega}} \nabla \mbox{d} \Omega +  X_{-}  \sigma^{-} + X_{+}  \sigma^{+} + 
\left[ \frac{1}{\Omega} \Delta \Omega -
\frac{3}{\Omega^{2}} \tilde{g}(\mbox{d} \Omega,\mbox{d} \Omega) \right] \tilde{g} \label{ricci} 
\end{equation}
where $\nabla=\nabla_{\sigma^{-}} + \nabla_{\sigma^{+}} $ is the connection of the product metric $\tilde{g}=\sigma^{-}+\sigma^{+}$.
\end{proposition}

Let us consider metrics with zero Cotton tensor again. If they have a constant argument, theorem \ref{th-cotton} implies that the principal structure is integrable and so, the spacetimes are of type D(M)$^{0 \cdot 0}_{0 \cdot 0}$. Consequently, from proposition \ref{pro-inte2} the Weyl tensor has real eigenvalues. So we can state:

\begin{theorem} \label{pro-wm}
The Weyl eigenvalues of a type D spacetime with zero Cotton tensor have constant argument if, and only if, they are real. 
\end{theorem} 
This result generalizes a previous one by McIntosh {\it et al.}\cite{mcar}. They have shown that there are no purely magnetic Type D vacuum metrics. But the purely magnetic case occurs when the eigenvalue argument is $\frac{3}{2} \theta =\pm \frac{\pi}{2}$, that is to say, a particular value of constant argument. So, from theorem \ref{pro-wm} it follows: 
 
\begin{corollary} \label{cor-wm}
There is no purely magnetic Type D metric with zero Cotton tensor.
\end{corollary}
This corollary shows that not only the purely magnetic vacuum solutions are forbidden, but also the Weyl-magnetic spacetimes  with zero Cotton tensor. On the other hand the McIntosh {\it et al.} result is also generalized in the sense that theorem \ref{pro-wm} excludes all the constant arguments that differ from $0$ or $\pi$.

From the results above it is easy to recover the canonical form for the metrics with zero Cotton tensor and real Weyl eigenvalues. Indeed, expressions (\ref{divcero1}) and (\ref{canonicfactor}) show that the conformal factor and the Weyl eigenvalue are related by $\Omega^2 = c^2 e^\rho = c^2 \alpha^{2/3}$, $c$ being an arbitrary constant. On the other hand they also satisfy expression (\ref{valorpropio}) and, consequently, $\Omega$ coincides, up to a constant factor, with $X_{-} +X_{+}$. So we have: 

\begin{proposition} \label{pro-canreal} 
Every type D metric with real eigenvalues and zero Cotton tensor writes 
$$g = \frac{1}{(X_{-} +X_{+} )^2} (\sigma^{-}+\sigma^{+}) $$
where  
 $\sigma^{-} = \sigma^{-}_{AB} (x^C) \mbox{d} x^A \mbox{d} x^B$,
$\sigma^{+} =\sigma^{+}_{ij}(x^k) \mbox{d} x^i \mbox{d} x^j $, are two arbitrary bidimensional metrics, 
$\sigma^{-}$ hyperbolic and $\sigma^{+}$ elliptic, with gaussian curvatures $X_{-} $ and $X_{+} $ respectively. 
\end{proposition} 
This canonical expression was obtained in a previous work\cite{fsS} where it was used to integrate the Einstein vacuum equations, in this way getting an intrinsic algorithm to identify every A, B and C-metric of Ehlers and Kundt\cite{ehku}. In the following section, starting from the propositions \ref{pro-inte1} and \ref{pro-inte2} we present a similar study for the charged counterpart of these vacuum solutions.

\section{Aligned Einstein-Maxwell solutions of type D$^{0 \cdot 0}_{0 \cdot 0}$}

If $(v,h)$ is the principal structure of the Weyl tensor, the aligned Einstein-Maxwell solutions satisfy 
\begin{equation} \label{ricci2} 
Ric(g) = \chi (v - h) = \kappa \ (\sigma^{-} - \sigma^{+} ) 
\end{equation} 
where the second equality is satisfied for the type D$^{0 \cdot 0}_{0 \cdot 0}$ metrics as a consequence of proposition \ref{pro-inte1}: $\chi = \kappa \Omega^2$, $\sigma^{-} = \Omega^2 v$, $\sigma^{+} = \Omega^2 h$. Moreover, as the principal structure is integrable, it is maxwellian and the associated Rainich index is a constant. So, (\ref{ricci2}) is a necessary and sufficient condition for the metric (\ref{canonica}) to be an aligned solution of the Einstein-Maxwell equations. Taking into account the expression (\ref{ricci}) for the Ricci tensor, condition (\ref{ricci2}) writes 
 \begin{eqnarray}
  \Omega=\lambda_{-}(x^A) + \lambda_{+}(x^i) \label{544} \\[3mm]
  \nabla \mbox{d} \lambda_{\epsilon}= \beta_{\epsilon} \ \sigma^{\epsilon} 
     \label{545} \\[3mm]
 \frac{\Omega^{2}}{6}(X_{-} + X_{+}) + \Omega \ (\beta_{-} + \beta_{+})= 
\sigma^{-}(\mbox{d}\lambda_{-},\mbox{d}\lambda_{-}) + \sigma^{+}(\mbox{d}\lambda_{+},\mbox{d} \lambda_{+}) \label{546} \\[3mm]
 \mbox{d} \beta_{\epsilon} + X_{\epsilon}  \mbox{d} \lambda_{\epsilon}=0  \label{547}
\end{eqnarray}
Equations (\ref{547}) are the integrability conditions of (\ref{545}). Moreover, if we differentiate (\ref{546}), project on $\sigma_-$, differentiate again and take into account (\ref{547}), we have
\begin{equation}
2(X_{-} + X_{+}) \mbox{d} \lambda_- \wedge \mbox{d} \lambda_+  + \Omega [ \mbox{d} X_+ \wedge \mbox{d} \lambda_-  -  \mbox{d} X_- \wedge \mbox{d} \lambda_+ ] = 0\   \label{548}
\end{equation}
Then a simple analysis of the expressions (\ref{544}-\ref{548}) leads to:

\begin{lemma} \label{lemma-sol}
The following conditions are equivalent: (i) $ \mbox{d} X_{\epsilon} = 0$, (ii) $\beta_{\epsilon} = 0$, (iii) $\mbox{d} \lambda_{\epsilon} = 0$,  (iv) $\sigma^{\epsilon}(\mbox{d} \Omega) = 0$. Moreover, these conditions hold if $\sigma^{\epsilon}(\mbox{d} \lambda_{\epsilon}, \mbox{d} \lambda_{\epsilon}) = 0$ everywhere.
\end{lemma}

\subsection{The solutions: A, B and C charged metrics}

Proposition \ref{pro-inte1} states that the classes D$^{0q0}_{0m0}$ can be discriminated using the vectors $\sigma^{\epsilon}(\mbox{d} \Omega)$. Then, lemma \ref{lemma-sol} implies that, as happens in the vacuum case\cite{fsS}, the four classes can be characterized by $\sigma^{-}$ or $\sigma^{+}$ to be bidimensional metrics that have constant curvature or not. 

If $g$ is in class D$^{010}_{010}$, lemma \ref{lemma-sol} implies that $\lambda_{\epsilon}$ can be taken as coordinate in the plane $\sigma^{\epsilon}$.  Then, equations (\ref{546}-\ref{548}) say that $\beta_{\epsilon}$, $X_{\epsilon} $ and $\sigma^{\epsilon}(d \lambda_{\epsilon} , d \lambda_{\epsilon})$ depend just on $\lambda_{\epsilon}$, and that 
$X_{\epsilon} = - \beta_{\epsilon}'$. Then, from (\ref{546}) we have $\beta_{-}'''(\lambda_{-}) + \beta_{+}'''(\lambda_{+}) = 0$ and, consequently, $\beta_{\epsilon}$ is a polynomial in $\lambda_{\epsilon}$ of degree less than or equal to three. But lemma \ref{lemma-sol} also states that $d \lambda_{\epsilon}$ is not a null vector everywhere. Then, Einstein-Maxwell equations (\ref{544}-\ref{548}) finally lead to:
\begin{equation}  \label{sigma1}
\sigma^{\epsilon} = \frac{1}{\epsilon f(\epsilon \lambda_{\epsilon})} \mbox{d}
\lambda_{\epsilon}^2 + f(\epsilon \lambda_{\epsilon}) dZ^2
\end{equation}
with $f(\lambda)$ a fourth degree polynomial. Then, putting (\ref{sigma1}) and (\ref{544}) in (\ref{canonica}) we recover the known expression of the charged C-metrics\cite{kra}.

If $g$ is in class D$^{000}_{010}$, lemma \ref{lemma-sol} implies that $\lambda_{-}$ can be taken as a coordinate in the plane $\sigma^{-}$ and, moreover, $\sigma^{+} $ must be of constant curvature. Thus, a redefinition of $\Omega$ and $\sigma^{-}$ allows us to consider $X_{+}  \in \{ -1 , 0 , 1 \}$ and $\Omega = \lambda_-$. Then, if we introduce the coordinate transformation $r= - \frac{1}{\lambda_-}$, a similar procedure that leads in the general case to the charged counterpart of the A$_i$-metrics: 
\begin{equation} \label{A-metric}
g=- a(r) \mbox{d}t^{2} + \frac{1}{a(r)}\mbox{d}r^{2} + r^2  \mbox{d}\sigma^2 \qquad \qquad  a(r) \equiv X - \frac{C}{r} + \frac{D}{r^2}
\end{equation}
$\mbox{d}\sigma^2$ being the bidimensional elliptic metric of constant curvature $X$, with $X = 1, -1, 0$ depending on the $A_1$, $A_2$ or $A_3$ case.

If $g$ is in the class D$^{010}_{000}$, in a similar way $\lambda_{+}$ can be taken as a coordinate in the plane $\sigma^{+}$, and $\sigma^{-} $ must be of constant curvature $X_{-}  \in \{ -1 , 0 , 1 \}$. Then, the coordinate transformation $r= - \frac{1}{\lambda_+}$, leads to the charged counterpart of the B$_i$-metrics: 
\begin{equation}
g= r^2  \mbox{d}\sigma^2 + a(r) \mbox{d}z^{2} + \frac{1}{a(r)}\mbox{d}r^{2}   \qquad \qquad  a(r) \equiv  X - \frac{C}{r} + \frac{D}{r^2}
\end{equation}
$\mbox{d}\sigma^2$ being the bidimensional hyperbolic metric of constant curvature $X$, with $X =1, -1, 0$ depending on the $B_1$, $B_2$ or $B_3$ case.

Finally, in class D$^{000}_{000}$ both bidimensional metrics have a constant curvature and equation (\ref{546}) implies that $X_{-} + X_{+} =0$. This means that the spacetimes is conformally flat and the metric writes $g=\sigma^{-}+\sigma^{+}$, where $\sigma^{\epsilon}$ are bidimensional metrics, hyperbolic and elliptic respectively, with a constant curvature $\epsilon X $. The metrics of this more degenerated class are the only ones that have zero Cotton tensor.

\subsection{The intrinsic characterization} 
The metrics of type D(M)$^{0 \cdot 0}_{0 \cdot 0}$, which take the canonical form (\ref{canonica}), admit an intrinsic identification by means of conditions involving the principal $2$--form ${\cal U}$. These characterization equations that we have given in previous sections could be written explicitly in terms of metric concomitants because ${\cal U}$ can be determined from the Weyl tensor\cite{fms}. Nevertheless, as a consequence of the Bianchi identities some of the above conditions can be satisfied identically taking into account the properties of the Ricci tensor. This is the case of vacuum metrics: as $Ric =0$ implies the nullity of the Cotton tensor, the principal planes always define an umbilical and maxwellian structure as a consequence of the results in section 4. Actually we want to characterize aligned Einstein-Maxwell solutions that are conformal to a product metric. So, the Weyl tensor must have real eigenvalues and the principal planes are the eigenspaces of the Ricci  tensor, that is, 
\begin{equation} \label{intrin1} 
W = 3 \alpha (U \otimes U - *U \otimes *U) + \alpha G , \qquad 
Ric = \chi (v-h)
\end{equation} 
where $ v = U^2$,  $h = - *U^2$. Then, taking into account the expressions in section 2 about 2+2 almost-product structures, a straightforward calculation shows that the Bianchi identities (\ref{bianchi1})
write 
\begin{eqnarray}
(3 \alpha + 2\chi ) \ Q_v = v \otimes h(d \alpha); \qquad 
(3 \alpha - 2\chi ) \ Q_h = h \otimes v(d \alpha) \label{intrin3} \\
v(d \chi ) - 2 \chi i(\delta U) U =0 ; \qquad 
h(d \chi ) + 2 \chi i(\delta *U ) *U=0 \label{intrin4} 
\end{eqnarray}
From these expressions we find that, under the scalar restriction $(3 \alpha)^2 \neq (2\chi)^2$, the properties of the structure follow just by imposing that the Weyl and the Ricci tensor take expressions (\ref{intrin1}). On the other hand, the case $(3 \alpha)^2 = (2\chi)^2$ leads to the {\it exceptional} metrics considered by Pleba\'{n}ski and Hacyan\cite{ple}. Nevertheless, it can easily be shown that $(3 \alpha)^2 \neq (2\chi)^2$ for the solutions recovered in the subsection above. So we get the following characterization:
\begin{lemma} \label{lemma-ABC}
The charged counterpart of the $A$, $B$ and $C$-metrics are the only aligned  
Einstein-Maxwell solutions of type D with real eigenvalues that satisfy $(3 \alpha)^2 \neq (2\chi)^2$, $\alpha$ and $\chi$ being, respectively, the Weyl and the Ricci eigenvalues. 
\end{lemma} 
Elsewhere\cite{fsS}, conditions for $g$ to be of type D with real eigenvalues have been given in terms of Weyl concomitants. In order to impose the Ricci tensor to take the form (\ref{intrin1}) we can use the algebraic Rainich conditions\cite{rai}. But if the Weyl tensor is of type D with real eigenvalues, a part of these Rainich conditions hold identically when we impose the aligned restriction. From these considerations and lemma \ref{lemma-ABC} we have: 
 
\begin{theorem} \label{propintrin1} 
The $A$, $B$ and $C$ Einstein-Maxwell solutions can be characterized by conditions 
   $$\alpha \neq 0 ; \qquad S^2 + S = 0 ; \quad Ric(x,x) \geq 0 $$
$$ {\rm \tr} Ric = 0, \qquad S[Ric] + Ric = 0;
\qquad (3 \alpha)^2 - (2 \chi)^2 \neq 0  $$
$W \equiv W(g)$ and $Ric \equiv  Ric(g)$ being the Weyl and Ricci tensors of the metric $g$, and where $\alpha = \alpha(g) \equiv - \left( \frac{1}{12} \ {\rm \tr} W^3 \right)^{\frac{1}{3}}$, $\chi = \chi(g)\ \equiv - \frac 12 ({\rm \tr}Ric^2)^{\frac 12}$,  $S = S(g) \equiv \frac{1}{3 \alpha} \left( W - \frac 12 \alpha g {\ppsim} g \right)$, $S[Ric]_{\alpha \beta} = S_{\alpha \mu \beta \nu} R^{\mu \nu}  $ and $x$ is an arbitrary time-like vector.
\end{theorem} 
This theorem offers an intrinsic and explicit description of the aligned Einstein-Maxwell solutions of type D$^{0 \cdot 0}_{0 \cdot 0}$. Now we look for an intrinsic and explicit way to identify every metric of this family, that is, to distinguish the $A_i$, $B_i$ and $C$ charged metrics. In a first step we must discriminate between the classes D$^{0p0}_{0m0}$ and, as a consequence of proposition \ref{pro-inte1}, this depends on the nullity of the vectors $v(d \Omega)$ and $h(d \Omega)$. But the expression (\ref{valorpropio}) for the Weyl eigenvalue and lemma \ref{lemma-sol} imply that, equivalently, the vectors $v(d \alpha)$ and $h(d \alpha)$ determine these properties. So, the same scheme as in the vacuum case\cite{fsS} can be used to distinguish between the classes. 

The last step to obtain the intrinsic and explicit characterization of the solutions is to get an invariant that provides the sign of the bidimensional curvature when this is constant. A straightforward calculation shows that if $X_{{\epsilon}} $  is constant, then
\begin{equation} \label{omega} 
X_{{\epsilon}}  \Omega^2 = \omega_{\epsilon} \equiv \frac{1}{9} (\mbox{d} \ln{(\alpha + \chi)})^2- 2 \alpha - \epsilon \chi
\end{equation}
So, we have a characterization of the Einstein-Maxwell  $A$, $B$ and $C$-metrics, and we recover the type D static vacuum solutions making $\chi=0$.
 
\begin{theorem} \label{propintrin2} 
Let $g$ be an aligned Einstein-Maxwell solution of type D$^{0\cdot0}_{0\cdot0}$ (characterized in theorem \ref{propintrin1}). Let us take the metric concomitants 
$$M \equiv *W ( d \alpha, \cdot , d \alpha, \cdot )  \hspace*{1cm} 
N \equiv S( d \alpha, \cdot , d \alpha, \cdot )$$
and let $x$ be an arbitrary unitary timelike vector. Then,
\begin{description}
\item (i) $g$ is a charged $C$-metric if, and only if, $M \neq 0$. 
\item (ii) $g$ is a charged $A$-metric if, and only if, $M=0$ and 
$2 N(x,x) + tr N > 0 $. 

Furthermore, it is of type $A_1$, $A_2$ or $A_3$ if 
$\omega_+ > 0$, $\omega_+ < 0$ or $\omega_+ =0$, respectively, where $\omega_+ \equiv \frac{1}{9} (\mbox{d} \ln{(\alpha + \chi)})^2- 2 \alpha - \chi$.
\item (iii) $g$ is a charged $B$-metric if, and only if, $M=0$ and 
$2 N(x,x) + tr N < 0 $. 

Furthermore, it is of type $B_1$, $B_2$ or $B_3$ if 
$\omega_- > 0$, $\omega_- < 0$ or $\omega_- =0$, respectively, where $\omega_- \equiv \frac{1}{9} (\mbox{d} \ln{(\alpha + \chi)})^2- 2 \alpha + \chi$.
\end{description} 
\end{theorem} 
This theorem provides an algorithm to identify, in the set of all metrics, the charged counterpart of the Ehlers and Kundt\cite{ehku} vacuum solutions. The particular case of the $A_1$ metrics corresponds to a charged spherically symmetric spacetime, that is, to the Reissner-Nordstr\"{o}m solution. In this case the metric takes the form (\ref{A-metric}) with $X = 1$, and the mass and the charge are related with the constants $C$ and $D$, respectively. Moreover, these constants can be given in terms of Weyl and Ricci invariants. Then, from the last theorem and previous subsection it follows:

\begin{theorem}
Let $Ric \equiv Ric(g)$ and $W\equiv W(g)$ be the Ricci and the Weyl tensors of a
spacetime metric $g$, and let us take the metric concomitants:
\begin{equation} \label{ro-chi} 
\hspace{-2.0cm}
\alpha \equiv -\left( \frac 1{12}{\rm \tr}W^3 \right)^{\frac 13} \qquad \chi \equiv - \frac 12 ({\rm \tr}Ric^2)^{\frac 12}  \qquad
\omega \equiv \frac 19 g(\mbox{d}ln\alpha ,\mbox{d}ln\alpha )-2\alpha - \chi 
\end{equation}
\begin{equation} \label{s-m-n}
 \hspace{-2.0cm} 
S\equiv \frac 1{3\alpha }(W- \frac{1}{2} \alpha \ g{\ppsim} g) 
\qquad M\equiv *W(\mbox{d}\alpha ,\cdot ,\mbox{d}\alpha ,\cdot )  \qquad  N \equiv S(
\mbox{d}\alpha ,\cdot ,\mbox{d}\alpha ,\cdot )
\end{equation}
The necessary and sufficient conditions for $g$ to be the Reissner-Nordstr\"{o}m
metric are
$$\alpha \neq 0 \qquad \qquad \quad S^2 + S = 0  \qquad \qquad Ric(x,x) \geq 0 $$
$$ {\rm \tr} Ric = 0 \quad \qquad S[Ric] + Ric = 0 \quad
\qquad (3 \alpha)^2 - (2 \chi)^2  \neq 0  $$
$$M=0 \quad \qquad 2N(x,x)+\mbox{tr}N>0\  \qquad \qquad \omega >0 \qquad$$
where $x$ is an arbitrary unitary time-like vector. Moreover, the
mass $m$ and the electric charge $e$ are given, respectively, by $m=\frac{ \alpha + \chi}{\omega ^{3/2}}$ and $e^2=-\frac{\chi}{\omega^2}$, and the time-like Killing vector by $\xi =\frac{1}{\sqrt{\omega}(3\alpha+2\chi)}\frac{N(x)}{\sqrt{N(x,x)}}$.
\end{theorem}

\subsection{A summary in algorithmic form}

Finally, in order to emphasize the algorithmic nature of our
results, we present them as a flow diagram that
identifies, among all metrics, every A, B or C Einstein-Maxwell solution (in the following flow chart we denote them $A^*$, $B^*$ and $C^*$-metrics). The exceptional metrics studied by Pleba\'nski are also identified and they are denoted $Exc$-metrics. This operational algorithm involves an arbitrary unitary timelike
vector, $x$, and some metric concomitants that may
be obtained from the components of the metric tensor $g$ in arbitrary
local coordinates: the invariants $\alpha$, $\chi$, $\omega_{\epsilon}$, $S$, $M$ and
$N$ are given in (\ref{omega}), (\ref{ro-chi}) and (\ref{s-m-n}) in terms of the Ricci and Weyl tensors. Making $\chi =0$, we recover the vacuum solutions\cite{fsS}.

\newpage
\setlength{\unitlength}{0.032cm}
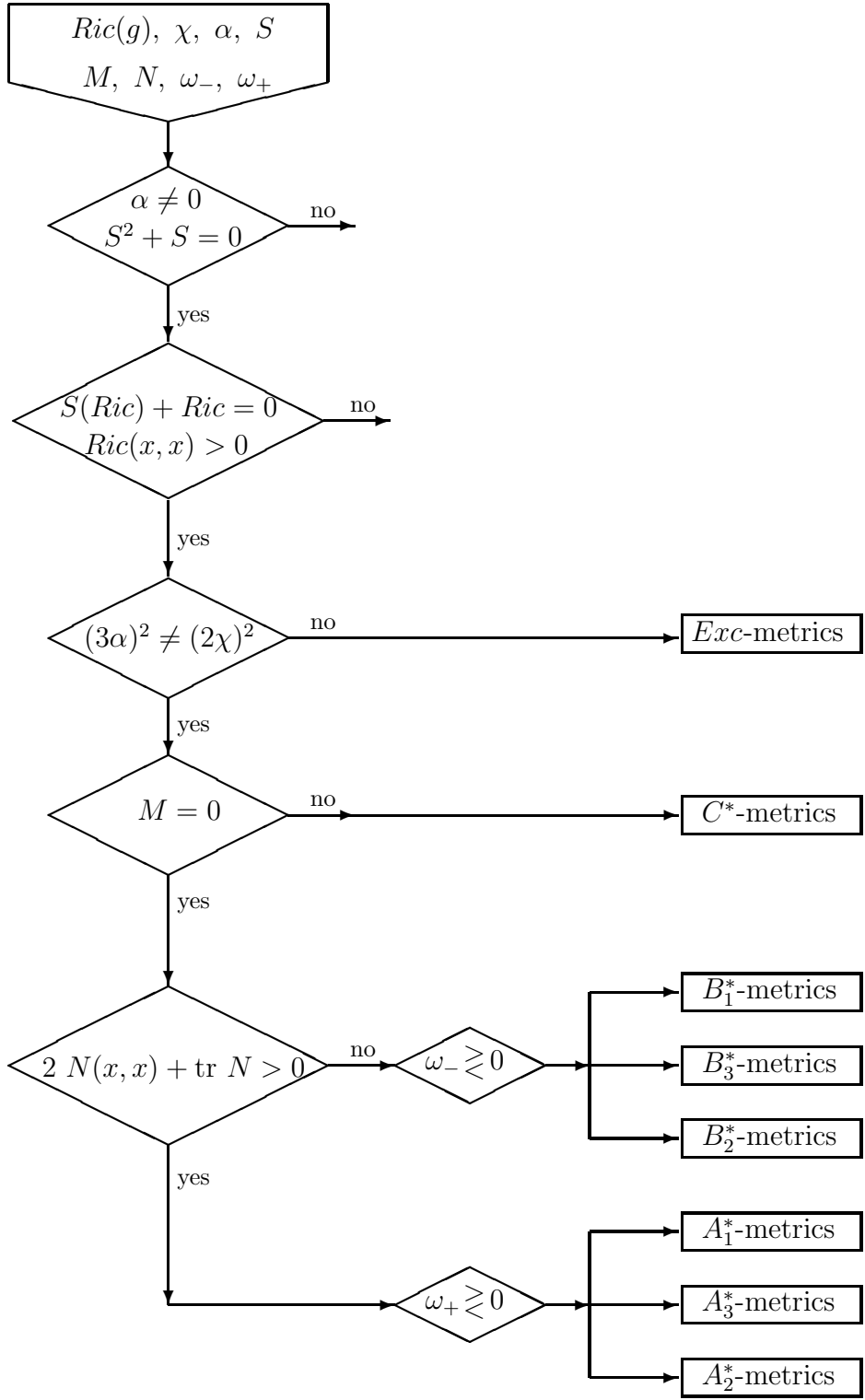
\begin{picture}(108,108)
\thicklines

\put(54,47){\vector(0,-1){20}}
\put(54,47){\line(4,1){72}}
\put(54,47){\line(-4,1){72}}
\put(126,65){\line(0,1){35}}
\put(-18,65){\line(0,1){35}}
\put(-18,100){\line(1,0){144}}
\put(10,85){$Ric(g) , \ \chi, \  \alpha,  \  S   $}
\put(15,63){$M,  \ N,  \ \omega_- , \ \omega_+$}

\put(0,0){\line(2,1){54}}
\put(0,0){\line(2,-1){54}}
\put(54,27){\line(2,-1){54}}
\put(54,-27){\line(2,1){54}}

\put(37,8){$\alpha \neq 0$}
\put(25,-9){$S^2 + S =0$}
\put(108,0){\vector(1,0){30}}
\put(118,4){\footnotesize{no}}
\put(58,-43){\footnotesize{yes}}

\put(-16,-88){\line(2,1){70}}
\put(-16,-88){\line(2,-1){70}}
\put(54,-53){\line(2,-1){70}}
\put(54,-123){\line(2,1){70}}
\put(54,-27){\vector(0,-1){25}}
\put(5,-86){$S( Ric)  +Ric=0$}
\put(16,-103){$ Ric(x,x)>0$}
\put(124,-88){\vector(1,0){30}}
\put(54,-124){\vector(0,-1){34}}
\put(136,-84){\footnotesize{no}}
\put(58,-144){\footnotesize{yes}}

\put(0,-186){\line(2,1){54}}
\put(0,-186){\line(2,-1){54}}
\put(54,-159){\line(2,-1){54}}
\put(54,-213){\line(2,1){54}}
\put(16,-190){$(3\alpha)^2 \neq  (2\chi)^2  $}
\put(108,-186){\vector(1,0){176}}
\put(286,-192){\framebox(80,16)[l]{$\;  Exc_{}$-metrics}}

\put(118,-182){\footnotesize{no}}
\put(54,-213){\vector(0,-1){26}}
\put(58,-228){\footnotesize{yes}}

\put(0,-266){\line(2,1){54}}
\put(0,-266){\line(2,-1){54}}
\put(54,-239){\line(2,-1){54}}
\put(54,-293){\line(2,1){54}}
\put(40,-268){$M=0$}
\put(108,-266){\vector(1,0){30}}
\put(118,-262){\footnotesize{no}}
\put(54,-293){\vector(0,-1){50}}
\put(58,-308){\footnotesize{yes}}

\put(-18,-379){\line(2,1){72}}
\put(-18,-379){\line(2,-1){72}}
\put(54,-343){\line(2,-1){72}}
\put(54,-415){\line(2,1){72}}
\put(-3,-384){$2 \ N(x,x) + \mbox{tr }N > 0 $}
\put(126,-379){\vector(1,0){30}}
\put(136,-375){\footnotesize{no}}
\put(54,-415){\vector(0,-1){72}}
\put(58,-433){\footnotesize{yes}}

\put(54,-487){\vector(1,0){102}}

\put(156,-379){\line(2,1){34}}
\put(156,-379){\line(2,-1){34}}
\put(190,-362){\line(2,-1){34}}
\put(190,-396){\line(2,1){34}}
\put(170,-381){$\omega_-$}
\put(187,-377){$>$}
\put(187,-385){$<$}
\put(199,-381){$0$}

\put(156,-487){\line(2,1){34}}
\put(156,-487){\line(2,-1){34}}
\put(190,-470){\line(2,-1){34}}
\put(190,-504){\line(2,1){34}}
\put(170,-489){$\omega_+$}
\put(187,-485){$>$}
\put(187,-493){$<$}
\put(199,-489){$0$}

\put(224,-379){\vector(1,0){20}}
\put(224,-487){\vector(1,0){20}}

\put(244,-412){\line(0,1){66}}
\put(244,-520){\line(0,1){66}}

\put(244,-346){\vector(1,0){40}}
\put(244,-379){\vector(1,0){40}}
\put(244,-412){\vector(1,0){40}}

\put(244,-487){\vector(1,0){40}}
\put(244,-454){\vector(1,0){40}}
\put(244,-520){\vector(1,0){40}}

\put(286,-354){\framebox(80,16)[l]{$\ \ B^{*}_{1}$-metrics}}
\put(286,-387){\framebox(80,16)[l]{$\ \ B^{*}_{3}$-metrics}}
\put(286,-420){\framebox(80,16)[l]{$\ \ B^{*}_{2}$-metrics}}

\put(286,-495){\framebox(80,16)[l]{$\ \ A^{*}_{3}$-metrics}}
\put(286,-462){\framebox(80,16)[l]{$\ \ A^{*}_{1}$-metrics}}
\put(286,-528){\framebox(80,16)[l]{$\ \ A^{*}_{2}$-metrics}}

\put(286,-274){\framebox(80,16)[l]{$\ \ C^{*}_{ }$-metrics}}

\put(138,-266){\vector(1,0){146}}

\end{picture}

\newpage

\section*{Acknowledgements}
The authors would like thank B. Coll and J. A. Morales for some useful comments.
This work has been supported by the Spanish Ministerio de Ciencia y Tecnolog\'{\i}a, project AYA2000-2045.

\end{document}

\end{document}